\documentclass[a4paper,12pt]{spieman}  
\usepackage{amsmath,amsfonts,amssymb}
\usepackage{graphicx}
\usepackage{setspace}
\usepackage{tocloft}
\usepackage{bbm}
\usepackage{bm}
\usepackage{nicefrac}
\usepackage{mathtools}
\usepackage{verbatim}
\usepackage{microtype} 
\usepackage{listings}
\usepackage{array}
\usepackage{wrapfig}
\usepackage{textcomp}
\usepackage{url}            
\usepackage{booktabs}       

\newcommand{\xbf}{\mathbf{x}}

\newcommand{\fbf}{\mathbf{f}}

\newcommand{\Pcal}{\mathcal{P}}
\newcommand{\Xbf}{\mathbf{X}}

\DeclareMathOperator{\prox}{\mathrm{prox}}
\newcommand{\etol}{\epsilon_{\mathrm{tol}}}

\newcommand{\NN}{\mathbb{N}}

\newcommand{\II}{\textbf{I}}
\newcommand{\CC}{\mathbb{C}}

\DeclareMathOperator*{\argmin}{arg\,min}

\newcommand{\F}{\mathcal{F}}
\newcommand{\J}{\mathcal{J}}
\newcommand{\K}{\mathcal{K}}
\newcommand{\G}{\mathcal{G}}
\newcommand{\R}{\mathcal{R}}

\newcommand{\T}{\mathcal{T}}

\newcommand{\N}{\mathcal{N}}

\newcommand{\M}{\mathcal{M}}

\newcommand{\W}{\mathcal{W}}
\newcommand{\HH}{\mathcal{H}}
\newcommand{\D}{\mathcal{D}}

\newcommand{\Scal}{\mathcal{S}}

\newcommand{\xt}{\mathbf{X}^{(t)}}

\newcommand{\xtp}{\mathbf{X}^{(t+1)}}

\newcommand{\epst}{\varepsilon_{t}}

\newcommand{\epstp}{\varepsilon_{t+1}}

\usepackage{algorithmic,algorithm}
\usepackage{tikz}

\title{A Brief Overview of Optimization-Based Algorithms for MRI Reconstruction Using Deep Learning }

\author[a]{Wanyu Bian}
\affil[a]{neuro42,  2 Bryant St. Suite 240, San Francisco, USA, 32601}

\cftpagenumbersoff{figure}
\cftpagenumbersoff{table} 
\begin{document} 
\maketitle

\begin{abstract}
Magnetic resonance imaging (MRI) is renowned for its exceptional soft tissue contrast and high spatial resolution, making it a pivotal tool in medical imaging. The integration of deep learning algorithms offers significant potential for optimizing MRI reconstruction processes. Despite the growing body of research in this area, a comprehensive survey of optimization-based deep learning models tailored for MRI reconstruction has yet to be conducted. This review addresses this gap by presenting a thorough examination of the latest optimization-based algorithms in deep learning specifically designed for MRI reconstruction. The goal of this paper is to provide researchers with a detailed understanding of these advancements, facilitating further innovation and application within the MRI community.

\end{abstract}


{\noindent \footnotesize\textbf{*}Wanyu Bian,  \linkable{wanyu@neuro42.com} }

\begin{spacing}{2}   

\section{Introduction}\label{sect:intro}  

Magnetic Resonance Imaging (MRI) is a crucial medical imaging technology that is non-invasive and non-ionizing, providing precise in-vivo images of tissues vital for disease diagnosis and medical research. As an indispensable instrument in both diagnostic medicine and clinical studies, MRI plays an essential role\cite{kim2000use,kasivisvanathan2018mri,bi2024exploration}.

Although MRI offers superior diagnostic capabilities, its lengthy imaging times, compared to other modalities, restrict patient throughput. This challenge has spurred innovations aimed at speeding up the MRI process, with the shared objective of significantly reducing scan duration while maintaining image quality\cite{pruessmann1999sense,griswold2002grappa}. Accelerating data acquisition during MRI scans is a major focus within the MRI and clinical application  community. Typically, scanning one sequence of MR images can take at least 30 minutes, depending on the body part being scanned, which is considerably longer than most other imaging techniques. However, certain groups such as infants, elderly individuals, and patients with serious diseases who cannot control their body movements, may find it difficult to remain still for the duration of the scan. Prolonged scanning can lead to patient discomfort and may introduce motion artifacts that compromise the quality of the MR images, reducing diagnostic accuracy. Consequently, reducing MRI scan times is crucial for enhancing image quality and patient experience.

MRI scan time is largely dependent on the number of phase encoding steps in the frequency domain (k-space), with common methods to accelerate the process involving the reduction of these steps by skipping phase encoding lines and sampling only partial k-space data. However, this approach can lead to aliasing artifacts due to under-sampling, violating the Nyquist criterion\cite{nyquist1928certain}. MRI reconstruction involves creating clear MR images from undersampled k-space data, which is then used for diagnostic and clinical purposes. Compressed Sensing (CS)\cite{donoho2006compressed} MRI reconstruction and parallel imaging\cite{sodickson1997simultaneous, pruessmann1999sense,larkman2007parallel} are effective techniques that address this inverse problem, speeding up MRI scans and reducing artifacts.

Deep learning has seen extensive application in image processing tasks \cite{siddique2021u, ronneberger2015u, zhou2018unet++, zhan2021deepmtl, zhan2022deepmtl} because of its ability to efficiently manage multi-scale data and learn hierarchical structures effectively, both of which are essential for precise image reconstruction and enhancement. Convolution neural network (CNN) also extensively utilized in MRI reconstruction due to its proficiency in handling complex patterns and noise inherent in MRI data\cite{hammernik2018learning,schlemper2017deep,han2018deep,zhu2018image,yang2017dagan,lee2018deep,knoll2020deep,liu2019santis,yaman2020self,blumenthal2022nlinv}. By learning from large datasets, deep learning algorithms can improve the accuracy and speed of reconstructing high-quality images, thus significantly enhancing the diagnostic capabilities of MRI technology.

\section{MRI Reconstruction Model}
Parallel imaging are k-space methods that utilize coil-by-coil auto-calibration, such as GRAPPA \cite{griswold2002generalized} and SPIRiT \cite{lustig2010spirit}. Compressed Sensing (CS)-based methods were applied on image domain such as SENSE \cite{pruessmann1999sense}, which depend on the accurate knowledge of coil sensitivity maps for optimization.
The formulation for the MRI reconstruction problem in CS-based parallel imaging is described by a regularized variational model as follows:
\begin{equation} \label{eq:ls}
    \min_x  \frac{1}{2} \|  Ax - f \|^2_2 + \mu R(x),
\end{equation}
where $x \in \CC^n$ is the MR image to be reconstructed, consisting of $n$ pixels, and $f \in \CC^m$ denotes the corresponding undersampled measurement data in k-space. $\mu > 0$ is a weight parameter that balance data fidelity term and regularization term. The measurement data is typically expressed as $f = Ax + \varepsilon$ with $\varepsilon \in \CC^m $ representing the noise encountered during acquisition.
The forward measurement encoding matrix $ A \in \CC^{m\times n}$ utilized in parallel imaging is defined by:
\begin{equation}\label{eq:encode}
    A := \Pcal_\Omega \F \Scal,
\end{equation}
where $\Scal := [\Scal_1, ..., \Scal_j]$ refers to the sensitivity maps of $ j$ different coils, $\F \in \CC^{n \times n}$ represents the 2D discrete Fourier transform, and $\Pcal_\Omega \in \NN^{m \times n}  (m \ll n)$  is the binary undersampling mask that captures $m$ sampled data points according to the undersampling pattern $\Omega$.

\section{Optimization-based network unrolling algorithms for MRI reconstruction}
Deep learning based model leverages large dataset and further explore the potential improvement of reconstruction performance comparing to traditional methods and has successful applications in clinic field\cite{knoll2020deep,hammernik2018learning,schlemper2017deep,han2018deep,zhu2018image,yang2017dagan,lee2018deep,liu2019santis,singh2023emerging,sun2020substituting,zhu2021deep}. 
Most existing deep-learning  based methods rendering end-to-end neural networks mapping from the partial k-space data to the reconstructed images \cite{WANG2020136,7493320,doi:10.1002/mp.12600,Quan2018CompressedSM,8417964}. To improve the interpretability of the relation between the topology of the deep model and reconstruction results, a new emerging class of deep learning-based methods known as \emph{learnable optimization algorithms} (LOA) have attracted much attention e.g. \cite{lundervold2019overview, liang2020deep, sandino2020compressed, mccann2017convolutional, zhou2020review, singha2021deep, chandra2021deep, ahishakiye2021survey,liu2020deep,bian2022optimal,jimaging7110231,bian2020DeepParallel,bian2020deep,bian2022MRIreconandSynth,bian2024relaxmore,bian2023magnetic,bian2024improving,bian2022learnable,bian2022optimization,bian2021optimization}. LOA was proposed to map existing optimization algorithms to structured networks where each phase of the networks correspond to one iteration of an optimization algorithm.

For instance, ADMM-Net \cite{NIPS2016_6406},  ISTA-Net$^+$ \cite{zhang2018ista}, and cascade network \cite{8067520} are regular MRI reconstruction.
Variational network (VN)\cite{doi:10.1002/mrm.26977} introduced gradient descent method by applying given sensitivities $ \Scal$. MoDL \cite{Aggarwal_2019} proposed a recursive network by unrolling the conjugate gradient algorithm using a weight sharing strategy. Blind-PMRI-Net \cite{10.1007/978-3-030-32251-9_80}  designed three network blocks to alternately update multi-channel images, sensitivity maps and the reconstructed MR image using an iterative algorithm based on half-quadratic splitting. The network in \cite{10.1007/978-3-030-32248-9_5} developed a Bayesian framework for joint MRI-PET reconstruction. VS-Net \cite{10.1007/978-3-030-32251-9_78} derived a variable splitting optimization method. However, existing methods still face the lack of accurate coil sensitivity maps and proper regularization in the parallel imaging problem. Alder et al.\cite{adler2018learned} proposed a reconstruction network that unrolled primal-dual algorithm where the proximal operator is learnable. DeepcomplexMRI \cite{WANG2020136} developed an end-to-end learning without explicitly using coil sensitivity maps to recover channel-wise images, and then combine to a single channel image in testing.

\subsection{Gradient Descent Algorithm Inspired Network}

\subsubsection{Variational Network}
Variational Network (VN) solves the model \eqref{eq:ls} by using gradient descent:
\begin{equation}\label{eq:vn}
    x^{(t+1)} = x^{(t)} - \alpha^{(t)} ( \lambda A^{\top} (Ax^{(t)} - f) + \nabla R(x^{(t)}) ).
\end{equation}
This model was applied on multi-coil MRI reconstruction. The regularization term was defined by Field of Expert model: $R(x) = \sum\limits_{i=1}^N <\HH_i (\G_i x ), \textbf{1} > $. A convolution neural network $\G_i$ is applied on the MRI data. The function $\HH_i$ is defined as nonlinear potential functions which is composed of scalar activation functions.  Then take the summation of the inner product of the non-linear term $\HH_i (\G_i x )$ and the vector of ones $\textbf{1}$. The sensitivity maps are pre-calculated and being used in $A$. The algorithm of VN unrolls the step \eqref{eq:vn} where the regularizer $R$ is parameterized by the learnable network $\G_i$ together with nonlinear activation function $ \HH_i$:
\begin{equation}\label{eq:vn-net}
     x^{(t+1)} = x^{(t)} - \alpha^{(t)} ( \lambda A^{\top} (Ax^{(t)} - f) + \sum_{i=1}^N (\G_{i}^{(t)})^{\top} \HH_i(\G_{i}^{(t)} x^{(t)})  ).
\end{equation}
 
\subsubsection{Variational model for joint reconstruction and synthesis}
This subsection introduces a provable learnable optimization algorithm\cite{bian2022MRIreconandSynth} for joint MRI reconstruction and synthesis.
Given the partial k-space data $\{ \fbf_1, \fbf_2 \}$ of the source modalities (e.g. T1 and T2), the goal of following model is to \textbf{reconstruct} the corresponding images $\{\xbf_1, \xbf_2\}$ \emph{and} \textbf{synthesize} the image $\xbf_3$ of the missing modality (e.g. FLAIR) without its k-space data:
\begin{equation}
\label{our_model_miccai}
\begin{aligned}
\min_{\xbf_1, \xbf_2, \xbf_3}\Psi_{\Theta, \gamma}(\xbf_1, \xbf_2, \xbf_3) & :=  \textstyle{\frac{1}{2} \sum\limits_{i = 1}^{2}  \| P_i F \xbf_i  - \fbf_i\|_2^2 + \frac{1}{3} \sum\limits_{i = 1}^{3}   \|h_{w_i} (\xbf_i) \|_{2,1}}  \\
&\quad + \textstyle{\frac{\gamma}{2} \|g_{\theta} ([h_{w_1}(\xbf_1),h_{ w_2} (\xbf_2)]) - \xbf_3\|_2^2}.
\end{aligned}
\end{equation}
This approach learns three modality-specific feature extraction operators $\{h_{w_i}\}_{i=1}^{3}$, one for each of these three modalities. 
The regularizers for three modalities are designed by combining these learned operators and a robust sparse feature selection operator ($(2,1)$-norm is used in this work).
To synthesize the image $\xbf_3$ using $\xbf_1$ and $\xbf_2$, another feature-fusion operator $g_{\theta}$ was employed which learns the mapping from the features $h_{w_1}(\xbf_1)$ and $h_{w_2}(\xbf_2)$ to the image $\xbf_3$.

Denote $\Xbf = \{\xbf_1, \xbf_2, \xbf_3 \}$, the forward Learnable Optimization Algorithm is presented in \eqref{alg:lda_joint}.
\begin{algorithm}[thb]
\caption{Learnable Descent Algorithm for joint MRI reconstruction and synthesis}
\label{alg:lda_joint}
\begin{algorithmic}[1]
\STATE \textbf{Input:} $\Xbf^{(0)}$, $0<\eta<1$, and $\varepsilon_0$, $a, \sigma >0$, $t = 0$. Max $T$, tolerance $\etol>0$.
\FOR{$t=0,1,2,\dots,T-1$}
\STATE $\xtp = \xt - \alpha_{t}  \nabla \Psi_{\Theta,\gamma}^{\epst} (\xt)$, where the step size $\alpha_{t}$ is obtained through \\ line search s.t. $ \Psi_{\Theta,\gamma}^{\epst}(\xtp) - \Psi_{\Theta,\gamma}^{\epst}(\xt) \le - \frac{1}{a} \| \xtp - \xt\|^2$ holds. \label{line_search}
\STATE \textbf{if} $\|\nabla \Psi_{\Theta,\gamma}^{\epst}(\xtp)\| < \sigma \eta {\epst}$,  set $\epstp= \eta {\epst}$;  \textbf{otherwise}, set $\epstp={\epst}$. \label{reduction_cre}
\STATE \textbf{if} $\sigma {\epst} < \etol$, 
\textbf{terminate} and go to Line \ref{endfor},
\ENDFOR{ and \textbf{output} $\Xbf^{(t)}$.} \label{endfor}
\end{algorithmic}
\end{algorithm}

In step 3, a gradient descent update applied with step size obtained by line search while the smoothing parameter $\epst > 0$ is fixed. In step 4, the reduction of $\epst$ ensures the subsequence who met the $\epst$ reduction criterion must have an accumulation point that is a Clarke stationary point of problem.

The backward network training algorithm is formulated to solve  a bilevel  optimization framework:
\begin{equation}
\label{eq:bi-level_miccai}
\begin{aligned}
  \min_{ \gamma}  \quad  \textstyle{\sum^{\mathcal{M}_{val}}_{i=1} }\ell( \Theta (\gamma) , \gamma ; \D^{val}_{i}) \quad 
 \mbox{s.t.}  \quad  \Theta(\gamma)  = \argmin_{\Theta} \textstyle{\sum^{\mathcal{M}_{tr}}_{i=1}} \ell ( \Theta , \gamma; \D^{tr}_{i}),
\end{aligned}
\end{equation}
\begin{equation}
\label{eq:def-l}
\begin{split}
 & \mbox{where}  \quad \ell ( \Theta , \gamma; \D_{i})  :=  \frac{\mu}{2} \|g_{\theta} ([h_{w_1}(\xbf_1^{*i}),h_{ w_2} (\xbf_2^{*i})]) - \xbf_3^{*i}\|_2^2 \\
 & + \textstyle{\sum_{j = 1}^{3}} \Big ( \frac{1}{2} \|\xbf_j^{(\hat{T})}({\Theta , \gamma}; \D_{i})  - \xbf_j^{*i} \|^2_2 
   + (1 - SSIM(\xbf_j^{(\hat{T})}({\Theta , \gamma}; \D_{i}), \xbf_j^{*i} ) ) \Big ) .
   \end{split}
\end{equation}
The following the penalty algorithm \eqref{alg:penalty} is proposed  to train the model for joint reconstruction and synthesis. 
\begin{algorithm}[htb]
\caption{Mini-batch alternating direction penalty algorithm}\label{alg:penalty}
\begin{algorithmic}[1]
\STATE \textbf{Input}  $\D^{tr}$, $\D^{val}$, $\delta_{tol}>0$, \textbf{Initialize}  $ \Theta$, $ {\gamma}$, $\delta$, $\lambda>0$ and $\nu_\delta \in(0, 1)$, \ $\nu_\lambda > 1$.
\WHILE{$\delta > \delta_{tol}$}
\STATE Sample training and validation batch $\mathcal{B}^{tr} \subset \D^{tr}, \mathcal{B}^{val} \subset \D^{val}$.
\WHILE{$\|\nabla_{\Theta}\widetilde{\mathcal{L}}( \Theta, \gamma ; \mathcal{B}^{tr}, \mathcal{B}^{val})\|^2 + \| \nabla_{\gamma}\widetilde{\mathcal{L}}( \Theta, \gamma ; \mathcal{B}^{tr}, \mathcal{B}^{val})\| ^2 > \delta$}
\FOR{$k=1,2,\dots,K$ (inner loop)}
\STATE $ \Theta \leftarrow \Theta - \rho_{\Theta}^{(k)} \nabla_{\Theta}\widetilde{\mathcal{L}}( \Theta, \gamma ; \mathcal{B}^{tr}, \mathcal{B}^{val})$
\ENDFOR
\STATE $ \gamma \leftarrow \gamma - \rho_{\gamma} \nabla_{\gamma}\widetilde{\mathcal{L}}( \Theta, \gamma ; \mathcal{B}^{tr}, \mathcal{B}^{val})$
\ENDWHILE{ and \textbf{update} $\delta \leftarrow \nu_\delta \delta$, $\ \lambda \leftarrow \nu_\lambda \lambda$. }
\ENDWHILE{ and \textbf{output:} $\Theta, {\gamma}$.}
\end{algorithmic}
\end{algorithm}

\subsection{Proximal Gradient Descent Algorithm Inspired Networks}
Solving inverse problem using proximal gradient descent has been largely explored and successfully applied in medical imaging reconstruction\cite{mardani2018neural,zeng2021review,bian2020DeepParallel,bian2020deep,bian2022optimal,bian2022MRIreconandSynth,bian2024relaxmore,bian2023magnetic,bian2024improving,bian2022learnable,bian2022optimization,bian2021optimization}.

Applying proximal gradient descent algorithm to approximate a (local) minimizer of \eqref{eq:ls} is an iterative process. The first step is gradient descent to force data consistency, and the second step applies a proximal operator to obtain updated image. The following steps iterates the proximal gradient descent algorithm:
\begin{subequations}\label{eq:prox}
\begin{align}
& b_{t} =  x_{t} - \rho_t A^{\top}   (A x_{t} - f), \label{eq:bi}  \\
& x_{t+1}  = \prox_{\rho_t R (\cdot)} (b_{t}), \label{eq:ui} \end{align}
\end{subequations}
where $\rho_t>0$ is the step size and $\prox_{\alpha R}$ is the proximal operator of $R$ defined by
\begin{equation}
    \prox_{\alpha R}(b) = \argmin_{x}  \frac{1}{2 \alpha} \| x - b \|_2^2 + R(x).
\end{equation}
The gradient update step \eqref{eq:bi} is straightforward to compute and fully utilizes the relation between the partial k-space data $f$ and the image $x$ to be reconstructed as derived from MRI physics.
This step involves implementing the proximal operation for regularization $\R$, which is equivalent to finding the maximum-a-posteriori solution for the Gaussian denoising problem at a noise level $\sqrt{\alpha}$ \cite{heide2014flexisp,venkatakrishnan2013plug}, thus the proximal operator can be interpreted as a Gaussian denoiser. However, because the proximal operator ${\rm prox}_{\mathcal{R}_\Theta,\rho_t}$ in the objective function \eqref{eq:ls} does not admit closed form solution, a CNN is used to substitute ${\rm prox}_{\mathcal{R}_\Theta,\rho_t}$, where the network can be constructed as a residual learning network\cite{mardani2018neural,bian2020DeepParallel,bian2022optimal,bian2024relaxmore} to avoid gradient vanishing problem. 

Mardani et al.\cite{mardani2018neural} introduced a recurrent neural network (RNN) architecture enhanced by residual learning to learn the proximal operator more effectively. This learnable proximal mapping  effectively functions as a denoiser, progressively eliminating aliasing artifacts from the input image.

\subsubsection{Parallel MRI Network}
Bian et al.\cite{bian2022optimal} developed a parallel MRI network leveraging residual learning to learn the proximal mapping and tackle model \eqref{eq:prox}, thus bypassing the requirement for pre-calculated coil sensitivity maps in the encoding matrix \eqref{eq:encode}. 
Parallel MRI network considers the MRI reconstruction problem as an bi-level optimization problem:
\begin{subequations}\label{learnable_pmri}
\begin{align}
& \min_{\Theta} \ \ \ell(  \xbf_{\Theta},  \xbf^*) ,\ \ \ \label{learnable_pmri_upper} \\
& \mathrm{s.t.} \ \  \xbf_{\Theta} = \argmin_{\xbf}  \phi_{\Theta}( \xbf). \ \ \label{learnable_pmri_lower} 
\end{align}
\end{subequations}
The variable $\xbf = (x_1,\dots, x_{c}) \in \mathbb{C}^{m \times n \times c}$ denotes the channel-specific multi-coil MRI data, with each $x_i$ corresponding to i-th coil for $i=1 \cdots c$. The study addresses a model  $\phi_{\Theta}$ that incorporates dual regularization terms applied to both image space and k-space, described by:
\begin{equation}\label{eq:m}
\phi(\xbf):= \frac{1}{2} \sum\limits^{c}_{i=1} \| \Pcal_\Omega \F  x_i -  f_i \|^2_2  +  R(\J(\xbf)) + R_f(\F x_i).
\end{equation}
The channel-combination operator $\J$ aims to learn a combination of multi-coil MRI data which integrate the prior information among multiple channels. Then the image domain regularizer $R$ extracts the information from the channel-combined image $\J(\xbf)$. The regularizer $R_f$ is designed to obtain prior information from k-space data. 

The upper level optimization \eqref{learnable_pmri_upper} is the network training process where the loss function $\ell(  \xbf_{\Theta},  \xbf^*) $ is defined as the discrepancy between learned $\xbf_{\Theta}$ and the ground truth $\xbf^*$.
The lower level optimization \eqref{learnable_pmri_lower} is solved by the following redefined algorithm:
\begin{subequations}\label{eq:newmodel} 
\begin{align}
b_i^{(t)}  = x_i^{(t)} - \rho_t  \F^{H}  \Pcal_\Omega^{\top}  (\Pcal_\Omega \F  x_i^{(t)} -  f_i),  & \quad i = 1,\cdots, c, \label{eq:newbi}  \\
\bar{x}_i^{(t)}   = [\prox_{\rho_t R(\J(\cdot)) } (b^{(t)})]_i, & \quad i = 1,\cdots, c, \label{eq:newubar} \\
x_i^{(t+1)}  = \prox_{ \rho_t R_f(\F(\cdot))}  (\bar{x}_i^{(t)}), & \quad i = 1,\cdots, c.\label{eq:newui}
\end{align}
\end{subequations}
The proximal operator can be understood as a Gaussian denoiser. Nevertheless, the proximal operator $ \prox_{\rho_t R}$ in the objective function \eqref{eq:newubar} lacks a closed-form solution, necessitating the use of a CNN as a substitute for $ \prox_{\rho_t R}$. This network is designed as a residual learning network denoted by $\phi$ in the image domain and $\varphi$ in the k-space domain, and the algorithm \eqref{eq:newmodel} is implemented in the following scheme:
\begin{subequations}  \label{eq:scheme}
\begin{align}
b_i{(t)} =  \ x_i{(t)} - \rho_t \F^{H}  \Pcal_\Omega^{\top} (\Pcal_\Omega \F x_i{(t)} - f_i),& \quad i = 1,\cdots, c, \label{eq:schemeb} \\
\bar{x}_i{(t)}  =  \ b_i{(t)} + \phi( b_i{(t)}), & \quad i = 1,\cdots, c, \label{eq:schemeu-bar}   \\
x_i{(t+1)} = \ \Bar{x}_i{(t)} +   \F^{H}  \varphi \big( \F (\Bar{x}_i{(t)} ) \big), & \quad i = 1,\cdots, c. \label{eq:schemeu} 
\end{align}
\end{subequations}
The CNN \( \phi \) utilizes channel-integration \( \J \) and operates with shared weights across iterations, effectively learning spatial features. However, it may erroneously enhance oscillatory artifacts as real features. In the k-space denoising step \eqref{eq:newui}, the k-space network \( \varphi \) focuses on low-frequency data, helping to remove high-frequency artifacts and restore image structure. Alternating between \eqref{eq:schemeu-bar}  and  \eqref{eq:schemeu} in their respective domains balances their strengths and weaknesses, improving overall performance.

This network architecture has also been generalized to quantitative MRI (qMRI) reconstruction problem under a self-supervised learning framework.

\subsubsection{RELAX-MORE}
RELAX-MORE \cite{bian2024relaxmore} introduced an optimization algorithm to unroll the proximal gradient for  qMRI reconstruction. RELAX-MORE is a self-supervised learning where the loss function minimizes the discrepancy between undersampled reconstructed MRI k-space data and the ``true'' undersampled k-space data retrospectively. The well-trained model can be applied to other testing data using transfer learning. As new techniques develop  \cite{pang2019transfer,zhuang2020comprehensive}, transfer learning may serve as an effective method to enhance the reconstruction timing efficiency of RELAX-MORE.

The qMRI reconstruction model aims to reconstruct the quantitative parameters $ \Psi $  and this problem can be formulated as a bi-level optimization model:
\begin{subequations} \label{bilevel}
\begin{align} 
 \min_{\Theta}  \,\,  & \ell( \Pcal_\Omega \F \Scal \M (\Psi \left( f \middle|\Theta\right) ), f ) \quad \text{s.t.} \label{up}\\
& \Psi( f | \Theta ) 
  = \argmin_{\Psi} \K_{\Theta} (\Psi), \label{low} \\
\text{where } \,\, & \K_{\Theta} \left(\Psi\right):=  \tfrac{1}{2}\parallel \Pcal_\Omega \F \Scal \M  \left(\Psi\right)- f \parallel_2^2+ \beta R_{\Theta} \left(\Psi\right). \label{relaxmore} 
\end{align}
\end{subequations}
The model $\M$ represents the MR signal function that maps the set of quantitative parameters $\Psi := \{ \psi_1, \cdots, \psi_N \}$ to the  MRI data. The loss function in \eqref{up} is addressed through a self-supervised learning network, and $\Psi ( f |\Theta ) $ is derived from the network parameterized by  $\Theta$. The upper level problem \eqref{up} focuses on optimizing the learnable parameters for network training, while the lower level problem \eqref{low} concentrates on optimizing the quantitative MR parameters.

Similar to the Parallel MRI Network\cite{bian2022optimal}, RELAX-MORE employs a proximal gradient descent algorithm to address the lower level problem \eqref{low}, with a residual network structure designed to learn the proximal mapping. Below is the unrolled learnable algorithm for resolving \eqref{low}:
 \begin{algorithm}
 \caption{Learnable Proximal Gradient Descent Algorithm }\label{alg:LDA}
 \begin{algorithmic}[1]
 \renewcommand{\algorithmicrequire}{\textbf{Input:}}
 \REQUIRE $\psi_i^{(0)}, \mu_i^{(1)}, \nu_i^{(1)},  \, i = 1, \cdots, N$. \label{input} 
\FOR {$t=1$  to $T$}  
\FOR {$i=1$  to $N$}  
\STATE $\bar{\psi}_i^{(t)}  = \psi_i^{(t-1)} -  \mu_i^{(t)} \nabla \frac{1}{2}  \| \Pcal_\Omega \F \Scal \M (\{\psi_i^{(t-1)} \}_{i=1}^N)  - f \|_2^2, $ \label{lda_gd}
\STATE $\psi_i^{(t)} = \widetilde{\W}_{\Theta_i}^{(t)} \circ \T_{\nu_i}^{(t)} \circ \W_{\Theta_i}^{(t)} (\bar{\psi}_i^{(t)}) + \bar{\psi}_i^{(t)} $ \label{lda_prox}
\ENDFOR
\STATE  $x^{(t)} = \M( \{\psi_i^{(t)} \}_{i=1}^N),\,\, $
\ENDFOR
\renewcommand{\algorithmicensure}{\textbf{Output:}}
\ENSURE  $ \{\psi_{i}^{(T)}\}_{i=1}^N$ \textbf{and} $ x^{(t)},   \forall t \in \{1, \cdots ,T\}$.  \label{lda_end}
 \end{algorithmic} 
 \end{algorithm}
 
Step \eqref{lda_prox} implements the residual network structure to learn the proximal operator with regularization $\beta R_{\Theta}$. The learnable operators $\widetilde{\W}_\Theta$ and $\W_\Theta$ has symmetric network structure, and $\T_{\nu}^{(t)}$ is the soft thresholding operator threshold parameter $\nu$.

\subsection{Alternating Direction Method of Multipliers (ADMM) Algorithm Inspired Networks}
ADMM considers to solve the following problem:
\begin{subequations}\label{eq:ls_sub}
\begin{align}
    \min\limits_{x,v} & \frac{1}{2}  \|  Ax - f \|^2_2 + \mu R(v).\\
    & \text{s.t.} \,\,\,  v - x = 0.
\end{align}
\end{subequations}
The ADMM algorithm solves the above problem by alternating the following three subproblems:
\begin{subequations}\label{eq:admm}
\begin{align}
x_{t+1} & = \argmin_x \| Ax - f \|^2_2 + \beta \| x - (v_{t} - u_{t}) \|_2^2 \\
v_{t+1} & = \beta \|  v_{t} - (x_{t+1} - u_{t}) \|_2^2 + \lambda R(v), \\
u_{t+1} & = u_{t} + (x_{t+1} - v_{t})
\end{align}
\end{subequations}

\subsubsection{ADMM-Net}
ADMM-Net\cite{NIPS2016_6406} reformulate these three steps through an augmented Lagrangian method. This approach leverages a cell-based architecture to optimize neural network operations for MRI image reconstruction. The network is structured into several layers, each corresponding to a specific operation in the ADMM optimization process. The Reconstruction layer  uses a combination of Fourier and penalized transformations to reconstruct images from undersampled k-space data, incorporating learnable penalty parameters and filter matrices. The Convolution layer  then applies a convolution operation, transforming the reconstructed image to enhance feature representation, using distinct, learnable filter matrices to increase the network's capacity. The Non-linear Transform layer replaces traditional regularization functions with a learnable piecewise linear function, allowing for more flexible and data-driven transformations that go beyond simple thresholding. Finally, the Multiplier Update layer updates the Lagrangian multipliers, essential for integrating constraints into the learning process, with learnable parameters to adaptively refine the model's accuracy. Each layer's output is methodically fed into the next, ensuring a coherent flow that mimics the iterative ADMM process, thus systematically refining the image reconstruction quality with each pass through the network.

\subsection{Primal-dual Hybrid Gradient (PDHG) Algorithm Inspired Networks}
PDHG can be used to solve the model \eqref{eq:ls}  by iterating the following steps:
\begin{subequations}\label{eq:pdhg}
\begin{align}
d_{t+1} & = \prox_{\zeta_t H^{\top}(\cdot)} ( d_{t} + \zeta_t A \bar{m}_{t}) \\
m_{t+1} & = \prox_{\rho_t R(\cdot)} ( m_t + \eta_t A^\top d_{t+1}), \\
\bar{m}_{t+1} & = m_{t+1} + \theta ( m_{t+1} -  m_{t}),
\end{align}
\end{subequations}
where $H$ is the function defined as $H(Ax, f) := \| Ax - f \|^2_2 $ in the model \eqref{eq:ls}. In the Learned PDHG\cite{mardani2018neural}, the traditional proximal operators are replaced with learned parametric operators. These operators are not necessarily proximal but are instead learned from training data, aiming to act similarly to denoising operators, such as Block Matching 3D (BM3D). The key innovation here is that these operators—both for the primal and dual  variables—are parameterized and optimized during training, allowing the model to learn optimal operation strategies directly from the data. The learned PDHG operates under a fixed number of iterations, which serves as a stopping criterion. This approach ensures that the computation time remains predictable and manageable, which is beneficial for time-sensitive applications. The algorithm maintains its structure but becomes more adaptive to specific data characteristics through the learning process, potentially enhancing reconstruction quality over traditional methods.

\subsection{Diffusion models meet gradient descent for MRI reconstruction }
A notable development for MRI reconstruction using diffusion model is the emergence of Denoising Diffusion Probabilistic Models (DDPMs)\cite{chung2022score,gungor2023adaptive,kazerouni2023diffusion,yang2023diffusion}. In Denoising Diffusion Probabilistic Models (DDPMs), the forward diffusion process systematically introduces noise into the input data, incrementally increasing the noise level until the data becomes pure Gaussian noise. This alteration progressively distorts the original data distribution. Conversely, the reverse diffusion process, or the denoising process, aims to reconstruct the original data structure from this noise-altered distribution. DDPMs effectively employ a Markov chain mechanism to transition from a noise-modified distribution back to the original data distribution via learned Gaussian transitions.  The learnable Gaussian noise can be parametrized in a U-net architecture that consists of transformers/attension layer\cite{vaswani2017attention} in each diffusion step. The Transformer model has demonstrated promising performance in generating global information and can be effectively utilized for image denoising tasks.\cite{zhang2023self}.

DDPMs represent an innovative class of generative models renowned for their ability to master complex data distributions and achieve high-quality sample generation without relying on adversarial training methods. Their adoption in MRI reconstruction has been met with growing enthusiasm due to their robustness, particularly in handling distribution shifts. Recent studies exploring DDPM-based MRI reconstructions \cite{chung2022score,gungor2023adaptive,kazerouni2023diffusion,yang2023diffusion} demonstrate how these models can generate noisy MR images which are progressively denoised through iterative learning at each diffusion step, either unconditionally or conditionally. This approach has shown promise in enhancing MRI workflows by speeding up the imaging process, improving patient comfort, and boosting clinical throughput. Moreover, several models \cite{chung2022score,chung2022come,gungor2023adaptive,guan2024correlated} have proven exceptionally robust, producing high-quality images even when faced with data that deviates from the training set (distribution shifts) \cite{knoll2019assessment}, accommodating various patient anatomies and conditions, and thus enhancing the accuracy and reliability of diagnostic imaging.

Chung et al. \cite{chung2022score} presented an innovative framework that applies score-based diffusion models to solve inverse imaging problems. The core technique involves training a continuous time-dependent score function using denoising score matching. During inference, the model alternates between a numerical Stochastic Differential Equation (SDE) solver and a data consistency step to reconstruct images. The method is agnostic to subsampling patterns, enabling its application across various sampling schemes and body parts not included in the training data. 

Bian et al. \cite{bian2023diffusion} proposed Domain-conditioned Diffusion Modeling (DiMo), which applies to both accelerated multi-coil MRI and quantitative MRI (qMRI) using diffusion models conditioned on the native data domain rather than the image domain. 
The method incorporates a gradient descent optimization within the diffusion steps to improve feature learning and denoising effectiveness. Here is the training and sampling algorithm for MRI recontruction:

 \begin{algorithm}
 \caption{Training Process of Static DiMo}\label{alg:static train}
 \begin{algorithmic}[1]
 \renewcommand{\algorithmicrequire}{\textbf{Input:}}
 \REQUIRE $ t \sim \text{Uniform}(\{ 1,\cdots, T\}), \epsilon \sim \N(\textbf{0}, \II), $ fully scanned k-space $\hat{f}_0 \sim q(f_0)  $, undersampling mask $\Pcal_\Omega$, partial scanned k-space $f$, and coil sensitivities $\Scal$.
 \\ \textit{Initialisation} : $\eta_0$
\STATE $ \hat{f}_t \leftarrow  \sqrt{\bar{\alpha}_t} \hat{f}_{0} + \sqrt{1- \bar{\alpha}_t }\epsilon.$ 
\STATE $ \hat{f}_t \leftarrow \Pcal_\Omega(\lambda_t f +(1-\lambda_t)\hat{f}_t ) + (\mathbbm{1} - \Pcal_\Omega) \hat{f}_t  $  
\hfill$\triangleright$ DC
\FOR {$k = 0$ to $K-1$}  
\STATE $ \hat{f}_t \leftarrow \hat{f}_t - \eta_k \nabla_{\hat{f}_t}  \frac{1}{2}\| A \F^{-1} \hat{f}_t - f\|^2_2 $ 
\hfill$\triangleright$ GD
\ENDFOR
\STATE Take gradient descent update step\\
\hspace{5pt} $ \nabla_{\theta} \| \epsilon  - \epsilon_{\theta} (\hat{f}_t ,t ) \|^2_2 $\\
\textbf{Until} converge\\
\renewcommand{\algorithmicensure}{\textbf{Output:}}
\ENSURE  $\hat{f}_t$, $t \in \{ 1,\cdots, T\}$. 
 \end{algorithmic} 
 \end{algorithm}
 \begin{algorithm}
 \caption{Sampling Process of Static DiMo}\label{alg:static sample}
 \begin{algorithmic}[1]
 \renewcommand{\algorithmicrequire}{\textbf{Input:}}
 \REQUIRE $ \hat{f}_T \sim \N(\textbf{0}, \II)$, undersampling mask $\Pcal_\Omega$, partial scanned k-space $f$, and coil sensitivities $\Scal$.\\
\FOR{$t=T-1,...,0$}
\STATE $z \sim \N(\textbf{0},\II)$ if  $t>0$, else $z=0$
\STATE $ \hat{f}_{t} = \mu_\theta(\hat{f}_{t+1},t+1 ) + \sigma_{t+1} z$

\STATE $ \hat{f}_{t}  \leftarrow \Pcal_\Omega(\lambda_{t}  f +(1-\lambda_{t} )\hat{f}_{t}  ) + (\mathbbm{1} - \Pcal_\Omega) \hat{f}_{t}   $  
\hfill$\triangleright$ DC
\FOR {$k = 0$ to $K-1$}
\STATE $ \hat{f}_{t}  \leftarrow \hat{f}_{t} - \eta_k \nabla_{\hat{f}_{t} }  \frac{1}{2}\| A \F^{-1} \hat{f}_{t} - f\|^2_2 $
\hfill$\triangleright$ GD
\ENDFOR
\ENDFOR
\renewcommand{\algorithmicensure}{\textbf{Output:}}
\ENSURE  $\hat{f}_0$ 
 \end{algorithmic} 
 \end{algorithm}
In the training and sampling algorithm, the gradient descent (GD) algorithm is applied iteratively into the diffusion step to refine k-space data further. It solves the optimization problem \eqref{eq:ls} without the regularization term. 

\section{Conclusion}
In conclusion, this paper provides a comprehensive overview of several optimization algorithms and network unrolling methods for MRI reconstruction. The discussed techniques include gradient descent algorithms, proximal gradient descent algorithms, ADMM, PDHG , and diffusion models combined with gradient descent. By summarizing these advanced methodologies, we aim to offer a valuable resource for researchers seeking to enhance MRI reconstruction through optimization-based deep learning approaches. The insights presented in this review are expected to facilitate further development and application of these algorithms in the field of medical imaging.


\appendix    

\bibliography{report}   

\begin{thebibliography}{10}

\bibitem{kim2000use}
R.~J. Kim, E.~Wu, A.~Rafael, {\em et~al.}, ``The use of contrast-enhanced magnetic resonance imaging to identify reversible myocardial dysfunction,'' {\em New England Journal of Medicine} {\bf 343}(20), 1445--1453  (2000).

\bibitem{kasivisvanathan2018mri}
V.~Kasivisvanathan, A.~S. Rannikko, M.~Borghi, {\em et~al.}, ``Mri-targeted or standard biopsy for prostate-cancer diagnosis,'' {\em New England Journal of Medicine} {\bf 378}(19), 1767--1777  (2018).

\bibitem{bi2024exploration}
Q.~Bi, X.~Lian, J.~Shen, {\em et~al.}, ``Exploration of radiotherapy strategy for brain metastasis patients with driver gene positivity in lung cancer,'' {\em Journal of Cancer} {\bf 15}(7), 1994  (2024).

\bibitem{pruessmann1999sense}
K.~P. Pruessmann, M.~Weiger, M.~B. Scheidegger, {\em et~al.}, ``Sense: Sensitivity encoding for fast mri,'' {\em Magnetic Resonance in Medicine} {\bf 42}(5), 952--962  (1999).

\bibitem{griswold2002grappa}
M.~A. Griswold, P.~M. Jakob, R.~M. Heidemann, {\em et~al.}, ``Generalized autocalibrating partially parallel acquisitions (grappa),'' {\em Magnetic Resonance in Medicine} {\bf 47}(6), 1202--1210  (2002).

\bibitem{nyquist1928certain}
H.~Nyquist, ``Certain topics in telegraph transmission theory,'' {\em Transactions of the American Institute of Electrical Engineers} {\bf 47}(2), 617--644  (1928).

\bibitem{donoho2006compressed}
D.~L. Donoho, ``Compressed sensing,'' {\em IEEE Transactions on information theory} {\bf 52}(4), 1289--1306  (2006).

\bibitem{sodickson1997simultaneous}
D.~K. Sodickson and W.~J. Manning, ``Simultaneous acquisition of spatial harmonics (smash): fast imaging with radiofrequency coil arrays,'' {\em Magnetic resonance in medicine} {\bf 38}(4), 591--603  (1997).

\bibitem{larkman2007parallel}
D.~J. Larkman and R.~G. Nunes, ``Parallel magnetic resonance imaging,'' {\em Physics in Medicine \& Biology} {\bf 52}(7), R15  (2007).

\bibitem{siddique2021u}
N.~Siddique, S.~Paheding, C.~P. Elkin, {\em et~al.}, ``U-net and its variants for medical image segmentation: A review of theory and applications,'' {\em Ieee Access} {\bf 9}, 82031--82057  (2021).

\bibitem{ronneberger2015u}
O.~Ronneberger, P.~Fischer, and T.~Brox, ``U-net: Convolutional networks for biomedical image segmentation,'' in {\em Medical image computing and computer-assisted intervention--MICCAI 2015: 18th international conference, Munich, Germany, October 5-9, 2015, proceedings, part III 18},  234--241, Springer  (2015).

\bibitem{zhou2018unet++}
Z.~Zhou, M.~M. Rahman~Siddiquee, N.~Tajbakhsh, {\em et~al.}, ``Unet++: A nested u-net architecture for medical image segmentation,'' in {\em Deep Learning in Medical Image Analysis and Multimodal Learning for Clinical Decision Support: 4th International Workshop, DLMIA 2018, and 8th International Workshop, ML-CDS 2018, Held in Conjunction with MICCAI 2018, Granada, Spain, September 20, 2018, Proceedings 4},  3--11, Springer  (2018).

\bibitem{zhan2021deepmtl}
C.~Zhan, M.~Ghaderibaneh, P.~Sahu, {\em et~al.}, ``Deepmtl: Deep learning based multiple transmitter localization,'' in {\em 2021 IEEE 22nd International Symposium on a World of Wireless, Mobile and Multimedia Networks (WoWMoM)},  41--50, IEEE  (2021).

\bibitem{zhan2022deepmtl}
C.~Zhan, M.~Ghaderibaneh, P.~Sahu, {\em et~al.}, ``Deepmtl pro: Deep learning based multiple transmitter localization and power estimation,'' {\em Pervasive and Mobile Computing} {\bf 82}, 101582  (2022).

\bibitem{hammernik2018learning}
K.~Hammernik, T.~Klatzer, E.~Kobler, {\em et~al.}, ``Learning a variational network for reconstruction of accelerated mri data,'' {\em Magnetic resonance in medicine} {\bf 79}(6), 3055--3071  (2018).

\bibitem{schlemper2017deep}
J.~Schlemper, J.~Caballero, J.~V. Hajnal, {\em et~al.}, ``A deep cascade of convolutional neural networks for mr image reconstruction,'' in {\em Information Processing in Medical Imaging: 25th International Conference, IPMI 2017, Boone, NC, USA, June 25-30, 2017, Proceedings 25},  647--658, Springer  (2017).

\bibitem{han2018deep}
Y.~Han, J.~Yoo, H.~H. Kim, {\em et~al.}, ``Deep learning with domain adaptation for accelerated projection-reconstruction mr,'' {\em Magnetic resonance in medicine} {\bf 80}(3), 1189--1205  (2018).

\bibitem{zhu2018image}
B.~Zhu, J.~Z. Liu, S.~F. Cauley, {\em et~al.}, ``Image reconstruction by domain-transform manifold learning,'' {\em Nature} {\bf 555}(7697), 487--492  (2018).

\bibitem{yang2017dagan}
G.~Yang, S.~Yu, H.~Dong, {\em et~al.}, ``Dagan: deep de-aliasing generative adversarial networks for fast compressed sensing mri reconstruction,'' {\em IEEE transactions on medical imaging} {\bf 37}(6), 1310--1321  (2017).

\bibitem{lee2018deep}
D.~Lee, J.~Yoo, S.~Tak, {\em et~al.}, ``Deep residual learning for accelerated mri using magnitude and phase networks,'' {\em IEEE Transactions on Biomedical Engineering} {\bf 65}(9), 1985--1995  (2018).

\bibitem{knoll2020deep}
F.~Knoll, K.~Hammernik, C.~Zhang, {\em et~al.}, ``Deep-learning methods for parallel magnetic resonance imaging reconstruction: A survey of the current approaches, trends, and issues,'' {\em IEEE signal processing magazine} {\bf 37}(1), 128--140  (2020).

\bibitem{liu2019santis}
F.~Liu, A.~Samsonov, L.~Chen, {\em et~al.}, ``Santis: sampling-augmented neural network with incoherent structure for mr image reconstruction,'' {\em Magnetic resonance in medicine} {\bf 82}(5), 1890--1904  (2019).

\bibitem{yaman2020self}
B.~Yaman, S.~A.~H. Hosseini, S.~Moeller, {\em et~al.}, ``Self-supervised learning of physics-guided reconstruction neural networks without fully sampled reference data,'' {\em Magnetic resonance in medicine} {\bf 84}(6), 3172--3191  (2020).

\bibitem{blumenthal2022nlinv}
M.~Blumenthal, G.~Luo, M.~Schilling, {\em et~al.}, ``Nlinv-net: Self-supervised end-2-end learning for reconstructing undersampled radial cardiac real-time data,'' in {\em ISMRM annual meeting},   (2022).

\bibitem{griswold2002generalized}
M.~A. Griswold {\em et~al.}, ``Generalized autocalibrating partially parallel acquisitions (grappa),'' {\em Magnetic Resonance in Medicine: An Official Journal of the International Society for Magnetic Resonance in Medicine} {\bf 47}(6), 1202--1210  (2002).

\bibitem{lustig2010spirit}
M.~Lustig and J.~M. Pauly, ``Spirit: iterative self-consistent parallel imaging reconstruction from arbitrary k-space,'' {\em Magnetic resonance in medicine} {\bf 64}(2), 457--471  (2010).

\bibitem{singh2023emerging}
D.~Singh, A.~Monga, H.~L. de~Moura, {\em et~al.}, ``Emerging trends in fast mri using deep-learning reconstruction on undersampled k-space data: a systematic review,'' {\em Bioengineering} {\bf 10}(9), 1012  (2023).

\bibitem{sun2020substituting}
H.~Sun, X.~Liu, X.~Feng, {\em et~al.}, ``Substituting gadolinium in brain mri using deepcontrast,'' in {\em 2020 IEEE 17th International Symposium on Biomedical Imaging (ISBI)},  908--912, IEEE  (2020).

\bibitem{zhu2021deep}
N.~Zhu, C.~Liu, X.~Feng, {\em et~al.}, ``Deep learning identifies neuroimaging signatures of alzheimer’s disease using structural and synthesized functional mri data,'' in {\em 2021 IEEE 18th International Symposium on Biomedical Imaging (ISBI)},  216--220, IEEE  (2021).

\bibitem{WANG2020136}
S.~Wang {\em et~al.}, ``Deepcomplexmri: Exploiting deep residual network for fast parallel mr imaging with complex convolution,'' {\em Magnetic Resonance Imaging} {\bf 68}, 136 -- 147  (2020).

\bibitem{7493320}
S.~{Wang}, Z.~{Su}, L.~{Ying}, {\em et~al.}, ``Accelerating magnetic resonance imaging via deep learning,'' in {\em 2016 IEEE 13th International Symposium on Biomedical Imaging (ISBI)},  514--517  (2016).

\bibitem{doi:10.1002/mp.12600}
K.~Kwon, D.~Kim, and H.~Park, ``A parallel mr imaging method using multilayer perceptron,'' {\em Medical Physics} {\bf 44}(12), 6209--6224  (2017).

\bibitem{Quan2018CompressedSM}
T.~M. Quan, T.~Nguyen-Duc, and W.-K. Jeong, ``Compressed sensing mri reconstruction using a generative adversarial network with a cyclic loss,'' {\em IEEE Transactions on Medical Imaging} {\bf 37}, 1488--1497  (2018).

\bibitem{8417964}
M.~{Mardani} {\em et~al.}, ``Deep generative adversarial neural networks for compressive sensing mri,'' {\em IEEE Transactions on Medical Imaging} {\bf 38}, 167--179  (2019).

\bibitem{lundervold2019overview}
A.~S. Lundervold and A.~Lundervold, ``An overview of deep learning in medical imaging focusing on mri,'' {\em Zeitschrift f{\"u}r Medizinische Physik} {\bf 29}(2), 102--127  (2019).

\bibitem{liang2020deep}
D.~Liang, J.~Cheng, Z.~Ke, {\em et~al.}, ``Deep magnetic resonance image reconstruction: Inverse problems meet neural networks,'' {\em IEEE Signal Processing Magazine} {\bf 37}(1), 141--151  (2020).

\bibitem{sandino2020compressed}
C.~M. Sandino {\em et~al.}, ``Compressed sensing: From research to clinical practice with deep neural networks: Shortening scan times for magnetic resonance imaging,'' {\em IEEE Signal Processing Magazine} {\bf 37}(1), 117--127  (2020).

\bibitem{mccann2017convolutional}
M.~T. McCann, K.~H. Jin, and M.~Unser, ``Convolutional neural networks for inverse problems in imaging: A review,'' {\em IEEE Signal Processing Magazine} {\bf 34}(6), 85--95  (2017).

\bibitem{zhou2020review}
S.~K. Zhou, H.~Greenspan, C.~Davatzikos, {\em et~al.}, ``A review of deep learning in medical imaging: Image traits, technology trends, case studies with progress highlights, and future promises,'' {\em Unknown Journal}   (2020).

\bibitem{singha2021deep}
A.~Singha, R.~S. Thakur, and T.~Patel, ``Deep learning applications in medical image analysis,'' {\em Biomedical Data Mining for Information Retrieval: Methodologies, Techniques and Applications} , 293--350  (2021).

\bibitem{chandra2021deep}
S.~S. Chandra, M.~Bran~Lorenzana, X.~Liu, {\em et~al.}, ``Deep learning in magnetic resonance image reconstruction,'' {\em Journal of Medical Imaging and Radiation Oncology}   (2021).

\bibitem{ahishakiye2021survey}
E.~Ahishakiye, M.~B. Van~Gijzen, J.~Tumwiine, {\em et~al.}, ``A survey on deep learning in medical image reconstruction,'' {\em Intelligent Medicine}   (2021).

\bibitem{liu2020deep}
R.~Liu, Y.~Zhang, S.~Cheng, {\em et~al.}, ``A deep framework assembling principled modules for cs-mri: Unrolling perspective, convergence behaviors, and practical modeling,'' {\em IEEE Transactions on Medical Imaging} {\bf 39}(12), 4150--4163  (2020).

\bibitem{bian2022optimal}
W.~Bian, Y.~Chen, and X.~Ye, ``An optimal control framework for joint-channel parallel mri reconstruction without coil sensitivities,'' {\em Magnetic Resonance Imaging}   (2022).

\bibitem{jimaging7110231}
W.~Bian, Y.~Chen, X.~Ye, {\em et~al.}, ``An optimization-based meta-learning model for mri reconstruction with diverse dataset,'' {\em Journal of Imaging} {\bf 7}(11)  (2021).

\bibitem{bian2020DeepParallel}
W.~Bian, Y.~Chen, and X.~Ye, ``Deep parallel mri reconstruction network without coil sensitivities,'' in {\em Machine Learning for Medical Image Reconstruction},  F.~Deeba, P.~Johnson, T.~W{\"u}rfl, {\em et~al.}, Eds., 17--26, Springer International Publishing, (Cham)  (2020).

\bibitem{bian2020deep}
W.~Bian, Y.~Chen, and X.~Ye, ``Deep parallel mri reconstruction network without coil sensitivities,'' in {\em Machine Learning for Medical Image Reconstruction: Third International Workshop, MLMIR 2020, Held in Conjunction with MICCAI 2020, Lima, Peru, October 8, 2020, Proceedings 3},  17--26, Springer  (2020).

\bibitem{bian2022MRIreconandSynth}
W.~Bian, Q.~Zhang, X.~Ye, {\em et~al.}, ``A learnable variational model for joint multimodal mri reconstruction and synthesis,'' in {\em Medical Image Computing and Computer Assisted Intervention -- MICCAI 2022},  L.~Wang, Q.~Dou, P.~T. Fletcher, {\em et~al.}, Eds., 354--364, Springer Nature Switzerland, (Cham)  (2022).

\bibitem{bian2024relaxmore}
W.~Bian, A.~Jang, and F.~Liu, ``Improving quantitative mri using self-supervised deep learning with model reinforcement: Demonstration for rapid t1 mapping,'' {\em Magnetic Resonance in Medicine}   (2024).

\bibitem{bian2023magnetic}
W.~Bian, A.~Jang, and F.~Liu, ``Magnetic resonance parameter mapping using self-supervised deep learning with model reinforcement,'' {\em ArXiv}   (2023).

\bibitem{bian2024improving}
W.~Bian, A.~Jang, and F.~Liu, ``Improving quantitative mri using self-supervised deep learning with model reinforcement: Demonstration for rapid t1 mapping,'' {\em Magnetic Resonance in Medicine}   (2024).

\bibitem{bian2022learnable}
W.~Bian, Q.~Zhang, X.~Ye, {\em et~al.}, ``A learnable variational model for joint multimodal mri reconstruction and synthesis,'' in {\em International Conference on Medical Image Computing and Computer-Assisted Intervention},  354--364, Springer  (2022).

\bibitem{bian2022optimization}
W.~Bian, {\em Optimization-Based Deep learning methods for Magnetic Resonance Imaging Reconstruction and Synthesis}.
\newblock PhD thesis, University of Florida  (2022).

\bibitem{bian2021optimization}
W.~Bian, Y.~Chen, X.~Ye, {\em et~al.}, ``An optimization-based meta-learning model for mri reconstruction with diverse dataset,'' {\em Journal of Imaging} {\bf 7}(11), 231  (2021).

\bibitem{NIPS2016_6406}
Y.~Yang {\em et~al.}, ``Deep admm-net for compressive sensing mri,'' in {\em Advances in Neural Information Processing Systems 29},  D.~D. Lee, M.~Sugiyama, U.~V. Luxburg, {\em et~al.}, Eds., 10--18, Curran Associates, Inc.  (2016).

\bibitem{zhang2018ista}
J.~Zhang and B.~Ghanem, ``Ista-net: Interpretable optimization-inspired deep network for image compressive sensing,'' in {\em Proceedings of the IEEE conference on computer vision and pattern recognition},  1828--1837  (2018).

\bibitem{8067520}
J.~{Schlemper} {\em et~al.}, ``A deep cascade of convolutional neural networks for dynamic mr image reconstruction,'' {\em IEEE Transactions on Medical Imaging} {\bf 37}(2), 491--503  (2018).

\bibitem{doi:10.1002/mrm.26977}
K.~Hammernik {\em et~al.}, ``Learning a variational network for reconstruction of accelerated mri data,'' {\em Magnetic Resonance in Medicine} {\bf 79}(6), 3055--3071  (2018).

\bibitem{Aggarwal_2019}
H.~K. Aggarwal, M.~P. Mani, and M.~Jacob, ``Modl: Model-based deep learning architecture for inverse problems,'' {\em IEEE Transactions on Medical Imaging} {\bf 38}, 394–405  (2019).

\bibitem{10.1007/978-3-030-32251-9_80}
N.~Meng {\em et~al.}, ``A prior learning network for joint image and sensitivity estimation in parallel mr imaging,'' in {\em Medical Image Computing and Computer Assisted Intervention -- MICCAI 2019},  732--740, Springer International Publishing, (Cham)  (2019).

\bibitem{10.1007/978-3-030-32248-9_5}
V.~P. Sudarshan {\em et~al.}, ``Joint reconstruction of pet + parallel-mri in a bayesian coupled-dictionary mrf framework,'' in {\em Medical Image Computing and Computer Assisted Intervention -- MICCAI 2019},  39--47, Springer International Publishing, (Cham)  (2019).

\bibitem{10.1007/978-3-030-32251-9_78}
J.~Duan {\em et~al.}, ``Vs-net: Variable splitting network for accelerated parallel mri reconstruction,'' in {\em Medical Image Computing and Computer Assisted Intervention -- MICCAI 2019},  713--722, Springer International Publishing, (Cham)  (2019).

\bibitem{adler2018learned}
J.~Adler and O.~{\"O}ktem, ``Learned primal-dual reconstruction,'' {\em IEEE transactions on medical imaging} {\bf 37}(6), 1322--1332  (2018).

\bibitem{mardani2018neural}
M.~Mardani, Q.~Sun, D.~Donoho, {\em et~al.}, ``Neural proximal gradient descent for compressive imaging,'' {\em Advances in Neural Information Processing Systems} {\bf 31}  (2018).

\bibitem{zeng2021review}
G.~Zeng, Y.~Guo, J.~Zhan, {\em et~al.}, ``A review on deep learning mri reconstruction without fully sampled k-space,'' {\em BMC Medical Imaging} {\bf 21}(1), 195  (2021).

\bibitem{heide2014flexisp}
F.~Heide, M.~Steinberger, Y.-T. Tsai, {\em et~al.}, ``Flexisp: A flexible camera image processing framework,'' {\em ACM Transactions on Graphics (ToG)} {\bf 33}(6), 1--13  (2014).

\bibitem{venkatakrishnan2013plug}
S.~V. Venkatakrishnan, C.~A. Bouman, and B.~Wohlberg, ``Plug-and-play priors for model based reconstruction,'' in {\em 2013 IEEE Global Conference on Signal and Information Processing},  945--948, IEEE  (2013).

\bibitem{pang2019transfer}
N.~Pang, L.~Qian, W.~Lyu, {\em et~al.}, ``Transfer learning for scientific data chain extraction in small chemical corpus with bert-crf model,'' {\em arXiv preprint arXiv:1905.05615}   (2019).

\bibitem{zhuang2020comprehensive}
F.~Zhuang, Z.~Qi, K.~Duan, {\em et~al.}, ``A comprehensive survey on transfer learning,'' {\em Proceedings of the IEEE} {\bf 109}(1), 43--76  (2020).

\bibitem{chung2022score}
H.~Chung and J.~C. Ye, ``Score-based diffusion models for accelerated mri,'' {\em Medical image analysis} {\bf 80}, 102479  (2022).

\bibitem{gungor2023adaptive}
A.~G{\"u}ng{\"o}r, S.~U. Dar, {\c{S}}.~{\"O}zt{\"u}rk, {\em et~al.}, ``Adaptive diffusion priors for accelerated mri reconstruction,'' {\em Medical Image Analysis} , 102872  (2023).

\bibitem{kazerouni2023diffusion}
A.~Kazerouni, E.~K. Aghdam, M.~Heidari, {\em et~al.}, ``Diffusion models in medical imaging: A comprehensive survey,'' {\em Medical Image Analysis} , 102846  (2023).

\bibitem{yang2023diffusion}
L.~Yang, Z.~Zhang, Y.~Song, {\em et~al.}, ``Diffusion models: A comprehensive survey of methods and applications,'' {\em ACM Computing Surveys} {\bf 56}(4), 1--39  (2023).

\bibitem{vaswani2017attention}
A.~Vaswani, N.~Shazeer, N.~Parmar, {\em et~al.}, ``Attention is all you need,'' {\em Advances in neural information processing systems} {\bf 30}  (2017).

\bibitem{zhang2023self}
D.~Zhang and F.~Zhou, ``Self-supervised image denoising for real-world images with context-aware transformer,'' {\em IEEE Access} {\bf 11}, 14340--14349  (2023).

\bibitem{chung2022come}
H.~Chung, B.~Sim, and J.~C. Ye, ``Come-closer-diffuse-faster: Accelerating conditional diffusion models for inverse problems through stochastic contraction,'' in {\em Proceedings of the IEEE/CVF Conference on Computer Vision and Pattern Recognition},  12413--12422  (2022).

\bibitem{guan2024correlated}
Y.~Guan, C.~Yu, Z.~Cui, {\em et~al.}, ``Correlated and multi-frequency diffusion modeling for highly under-sampled mri reconstruction,'' {\em IEEE Transactions on Medical Imaging}   (2024).

\bibitem{knoll2019assessment}
F.~Knoll, K.~Hammernik, E.~Kobler, {\em et~al.}, ``Assessment of the generalization of learned image reconstruction and the potential for transfer learning,'' {\em Magnetic resonance in medicine} {\bf 81}(1), 116--128  (2019).

\bibitem{bian2023diffusion}
W.~Bian, A.~Jang, and F.~Liu, ``Diffusion modeling with domain-conditioned prior guidance for accelerated mri and qmri reconstruction,'' {\em arXiv preprint arXiv:2309.00783}   (2023).

\end{thebibliography}
\bibliographystyle{spiejour}   


\vspace{2ex}\noindent\textbf{Wanyu Bian} is an Image Reconstruction Scientist at Neuro42, specializing in mathematical problems of image and signal processing as well as MRI reconstruction using a blend of optimization and deep learning techniques. In this role, she has pioneered the design of a magnet array optimization algorithm, advancing the field significantly. Previously, she enhanced her expertise as a Research Fellow at Harvard Medical School, focusing on complex challenges in medical imaging. Her academic foundation includes a PhD in Applied Mathematics from the University of Florida, which provided her with deep insights into machine learning, signal processing, MRI, and quantitative imaging.

\end{spacing}
\end{document}